\documentclass[twocolumn]{webofc}
\usepackage[varg]{txfonts}
\usepackage{hyperref}

\begin{document}
\title{Unbound states in $\mathbf{^{19}}$O(d,p$\boldsymbol{\gamma}$)$\mathbf{^{20}}$O: tracing the $\boldsymbol{\nu}\mathbf{(d_{3/2})}$ orbital}

\author{
    \firstname{Charlie James} \lastname{Paxman}\inst{1}\fnsep\thanks{\email{charlie.paxman@ganil.fr}} 
\and
    \firstname{Irene} \lastname{Zanon}\inst{2}
\and
    \firstname{Emmanuel} \lastname{Clément}\inst{1}
\and
    \firstname{Alain} \lastname{Goasduff}\inst{3}
\and
    \firstname{Javier} \lastname{Menendez}\inst{4}
\and
    \firstname{Takayuki} \lastname{Miyagi}\inst{5,6,7}
\and
    \firstname{Marlene} \lastname{Assié}\inst{8}
\and
    \firstname{Michal} \lastname{Ciemala}\inst{9}
\and
    \firstname{Freddy} \lastname{Flavigny}\inst{10}
\and
    \firstname{Antoine} \lastname{Lemasson}\inst{1}
\and
    \firstname{Adrien} \lastname{Matta}\inst{10}
\and
    \firstname{Diego} \lastname{Ramos}\inst{1}
\and
    \firstname{Mauricy} \lastname{Rejmund}\inst{1}
\and   
    \lastname{the \textit{e775s} Collaboration}
}

\institute{
    Grand Acc\'{e}l\'{e}rateur National d’Ions Lourds (GANIL), CEA/DRF-CNRS/IN2P3, Bvd Henri Becquerel, 14076 Caen, France 
\and
    KTH - Royal Institute of Technology, Stockolm, SE-10044, Sweden
\and
    INFN Laboratori Nazionali di Legnaro, Legnaro, Italy
\and    
    Department of Quantum Physics and Astrophysics and Institute of Cosmos Sciences, University of Barcelona, Spain
\and
    Department of Physics, Technische Universität Darmstadt, Darmstadt, Germany
\and
    ExtreMe Matter Institute, GSI Helmholtzzentrum für Schwerionenforschung GmbH, Darmstadt, Germany
\and
    Max-Planck-Institut für Kernphysik, Heidelberg, Germany
\and
    Université Paris-Saclay, CNRS/IN2P3, IJCLab, 91405 Orsay, France
\and
    IFJ PAN, Krakow, Poland
\and
    Université de Caen Normandie, ENSICAEN, CNRS/IN2P3, LPC Caen UMR6534, F-14000 Caen, France
          }

\abstract{%
  The single-neutron transfer reaction $^{19}$O(d,p$\gamma$)$^{20}$O has been performed at GANIL, populating states up to and above the neutron separation energy. Bound states populated by $s$-wave and $d$-wave transfer have been observed with improved experimental angular distributions. Critically, several unbound states between 7.6 MeV and 9.8 MeV have been accessed for the first time through the $^{19}$O(d,p) channel, and isolated via particle-$\gamma$ spectroscopy.
}
\maketitle
\section{Introduction}
\label{intro}

The neutron dripline in oxygen isotopes presents a clear challenge and unique opportunity for studies of shell evolution and nuclear structure. The heaviest observed bound isotope of flourine ($Z=9$) has 22 neutrons, whereas oxygen -- with only one fewer proton, $Z=8$ -- can only bind 16 neutrons. This striking anomaly is a result of an increase in the spacing between the $\nu(d_{3/2})$ orbital and the $\nu(s_{1/2} d_{5/2})$ orbitals, which was only explained by the inclusion of three-body forces~\cite{Otsuka2010}. As such, measurements relating to the $\nu(d_{3/2})$ orbital in oxygen isotopes are of significant interest as tests of modern theoretical models. We here present preliminary results of the first analysis of unbound states populated in the $^{19}$O(d,p$\gamma$)$^{20}$O reaction, in search of states originating from the high-lying $\nu(d_{3/2})$ orbital.

\section{Experimental method}
\label{method}

This work is a reanalysis of the work of I.~Zanon \textit{et al.}~\cite{Zanon2023_PRL} wherein a radioactive beam of $^{19}$O was delivered by the GANIL-SPIRAL1+ accelerator complex at 8 MeV/u and with an average intensity of $4\times10^{5}$ particles per second. This beam then impinged on thin CD$_{2}$ targets, alternately with and without a gold foil backing. Following the $^{19}$O(d,p) direct reaction, the $^{20}$O heavy recoil nucleus was detected at the focal plane of the VAMOS++ magnetic spectrometer~\cite{Rejmund2011_VAMOS}, the proton ejectile was detected in the trapezoidal detectors of the MUGAST array~\cite{Assie2021_MAVcampaign}, and the $\gamma$-ray de-excitations were observed in the high-resolution AGATA tracking array~\cite{Akkoyun2012_AGATA}. The energy and angle of the detected protons was used to kinematically reconstruct the excitation energy of the $^{20}$O nucleus. In addition, Doppler correction of $\gamma$-rays was performed event-by-event using precise kinematic information from the detected proton.

While I.~Zanon's work was focused on extracting femtosecond lifetimes using the AGATA array, using the MUGAST detector for channel selection, we here turn our attention to the particle spectroscopy, using the coincident $\gamma$-rays to isolate states. This plurality of simultaneous results, each with different goals and experimental focuses, emphasises the rich data sets that are achievable with the unique coupling of MUGAST-AGATA-VAMOS++.~\cite{Assie2021_MAVcampaign}

\section{Preliminary results}
\label{results}

\begin{figure*}[t]
\centering
\includegraphics[width=0.9\linewidth]{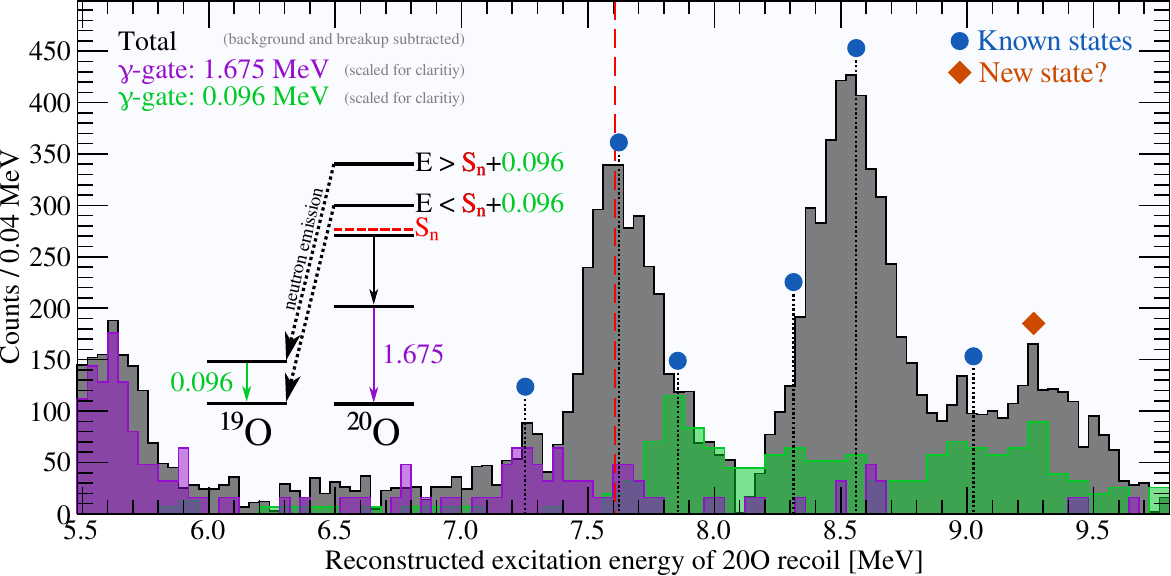}
\caption{High-energy and unbound region of the $^{19}$O(d,p$\gamma$)$^{20}$O preliminary spectrum. The full spectrum is shown in grey, with the spectra in coincidence with the 0.096~MeV (1.675~MeV) $\gamma$-ray shown in purple (green). The $\gamma$-gated spectra have been arbitrarily scaled for clarity.}
\label{fig_unbgammagates}
\end{figure*}

In Fig.~\ref{fig_unbgammagates}, we present the excitation energy spectrum of $^{20}$O, focusing here only on the high-energy region, between 5.5~MeV and 9.8~MeV. The upper energy limit is enforced by the energy threshold of the MUGAST detectors. The background-subtracted excitation spectrum is shown in grey, overlayed with two $\gamma$-gated excitation spectra, observed in coincidence with the 1.675~MeV, $2_1^++\rightarrow0_{g.s.}^+$ transition in $^{20}$O (in purple), and the 0.096~MeV, $3/2_1^++\rightarrow5/2_{g.s.}^+$ transition in $^{19}$O (in green). Critically, the observation of a $^{19}$O $\gamma$-ray in coincidence with the unbound region of $^{20}$O indicates the population of narrow unbound states (see inlay of Fig.~\ref{fig_unbgammagates}).

Several known states~\cite{LaFrance1979_18Otp20O,Bohlen2011_14C7Lip20O,Sumithrarachchi2006_20Nbeta20O} are marked in Fig.~\ref{fig_unbgammagates} -- namely 7.25~MeV, 7.62~MeV, 7.75~MeV, 8.31~MeV, 8.56~MeV and 9.03~MeV -- and appear in this preliminary analysis to be populated here for the first time via $^{19}$O(d,p) transfer. The two strongest states, 7.62~MeV and 8.56~MeV, appear to not decay significantly via the excited $3/2_1^+$ $^{19}$O state. For 7.62~MeV, this is explained simply by energy matching arguments; the state is less than 96~keV over the neutron separation threshold, and therefore could not decay through this route, and would instead proceed directly to the ground state of $^{19}$O. For 8.56~MeV, the reasoning is less clear, as it could decay to either the first or second excited states of $^{19}$O by energy matching arguments. The apparent small branching ratio to the 96~keV state may suggest a favourable single-particle structure overlap between this state and the $\nu(d_{5/2}^3)$ $^{19}$O ground state. Conversely, the excited $3/2_1^+$ $^{19}$O state has a $\nu(d_{5/2}^2d_{3/2}^{1})$ structure, which may inform the structure of unbound $^{20}$O states which decay via this route; namely, the 7.62~MeV, 7.75~MeV, 9.03~MeV states and a potential novel state, which appears at 9.3~MeV.

\section{Future work}

An analysis of bound and unbound differential cross sections is in progress, and theoretical interpretation of these experimental cross-sections is underway. Following the extraction of spectroscopic factors, and we aim to compare our experimental observations to shell model calculations to inform our understanding of the $\nu(d_{3/2})$ orbital evolution in neutron-rich oxygen as we move towards the dripline.

\vspace{40pt}%

\end{document}